\documentclass[aps,prb,twocolumn,showpacs,groupedaddress]{revtex4}  
\usepackage{graphicx}  
\usepackage{amssymb}   
\usepackage{amsmath}
\usepackage{amsfonts}
\usepackage{amstext}

\begin{document}

\title{Phase shifts and phase $\pi$-jumps in four-terminal waveguide Aharonov-Bohm interferometers}
\date{\today}

\author{Christoph Kreisbeck}
\author{Tobias Kramer}\email[electronic address: ]{tobias.kramer@physik.uni-regensburg.de}
\affiliation{Institut f\"ur theoretische Physik, Universit\"at Regensburg, Germany}
\author{Sven S.\ Buchholz}
\author{Saskia F.\ Fischer}\email[electronic address: ]{saskia.fischer@rub.de}
\author{Ulrich Kunze}
\affiliation{Lehrstuhl f\"ur Werkstoffe und Nanoelektronik, Ruhr-Universit\"at Bochum, 44780 Bochum, Germany}
\author{Dirk Reuter}
\author{Andreas D.\ Wieck}
\affiliation{Lehrstuhl f\"ur Angewandte Festk\"orperphysik, Ruhr-Universit\"at Bochum, 44780 Bochum, Germany}

\begin{abstract}
Quantum coherent properties of electrons can be studied in Aharonov-Bohm (AB) interferometers. We investigate both experimentally and theoretically the transmission phase evolution in a four-terminal quasi-one-dimensional AlGaAs/GaAs-based waveguide AB ring. As main control parameter besides the magnetic field, we tune the Fermi wave number along the pathways using a top-gate. Our experimental results and theoretical calculations demonstrate the strong influence of the measurement configuration upon the AB-resistance-oscillation phase in a four-terminal device. While the non-local setup displays continuous phase shifts of the AB oscillations, the phase remains rigid in the local voltage-probe setup. Abrupt phase jumps are found in all measurement configurations. We analyze the phase shifts as functions of the magnetic field and the Fermi energy and provide a detailed theoretical model of the device. Scattering and reflections in the arms of the ring are the source of abrupt phase jumps by $\pi$.
\end{abstract}

\pacs{73.21.Hb, 73.23.Ad, 85.35.Ds}
\maketitle 

\section{Introduction}\label{sec:introduction}

The magnitude and phase of electron-wave transmission are of high interest for fundamental investigations in solid-state quantum devices and circuits. AB interferometers have been used as probes to study coherence properties of systems embedded in one of the interferometer arms, such as a quantum dot (QD).\cite{yac95,schu97,koba0204a, koba0204b,sigr04,ji00a,ji00b,ji00c,ji00d,kats07} These experiments showed unexpected features like abrupt phase jumps by $\pi$ raising the question how exact such a phase determination can be.\cite{Oreg1997a,enti02,Silva2002a} In quantum rings fabricated from quasi-one dimensional (1D) quantum waveguides the impact of scattering and reflection of electron waves, e.g. at cross-junctions and leads, on the magnitude and phase of transmission under realistic measurement and circuitry conditions remain yet unresolved. Here, we present a comprehensive investigation including the detailed comparison of experimental results and realistic theoretical modeling of a ring device which allows for the detection of an intrinsic (electrostatic) transmission phase shift.

The AB effect allows one to detect interference of coherent electrons in a two-path ring in the form of magnetoresistance oscillations with a magnetic flux period of $h/e$.\cite{ahar59,webb85} If the lengths of the two paths $s_1$ and $s_2$ differ, an additional wave-number-dependent phase occurs, given by $\Delta\alpha=k_\mathrm{F}(s_2-s_1)$. Ideally, the transmission probability along the paths becomes $T\propto \mathrm{cos}(e\phi/\hbar+\Delta\alpha)$, with magnetic flux $\phi$. The wave number $k_\mathrm{F}$ can be controlled by a perpendicular electric field applied via a top-gate electrode, which might only cover part of the device.\cite{koba02}

The simple linear relation between wave-number and phase $\Delta\alpha$ does not take into account time-reversal symmetry, which enforces $T(\phi)=T(-\phi)$ in two-terminal devices \cite{onsa31,casi45,butt88} and thus no continuous phase shifts can be detected.\cite{schu97,Cernicchiaro1997,Pedersen1999a} In order to break the phase rigidity, it is necessary to reduce the device symmetry by attaching additional leads to the ring. \cite{butt88} The addition of leads increases scattering effects in the cross-junctions and requires to model the device in a two-dimensional fashion.

In Sect.~\ref{sec:experiment} we describe our asymmetric four-terminal quasi-1D waveguide interferometer with orthogonal cross-junctions and discuss the experimental results.

Sect.~\ref{sec:theory} contains the theoretical two-dimensional device model, which goes beyond effective 1D models.\cite{Gefen1984a,Buttiker1984a,Vasilopoulos2007a,Hedin2009a,Ying2008a} The theoretical calculations encompass a large range of Fermi energies and are efficiently performed using the wave-packet approach to mesoscopic transport.\cite{Kramer2010b,Kramer2008a} The inclusion of non-zero bias-voltages and temperature allows us to compare experiments and theory on an unprecedented level of detail.

In Sect.~\ref{sec:comparison} we relate the occurrence of abrupt phase jumps, which have been observed in nearly all AB experiments, to resonances forming due to multiple reflections along the ring paths.

\begin{figure}[t]
\begin{center}
\includegraphics[height=0.55\textheight]{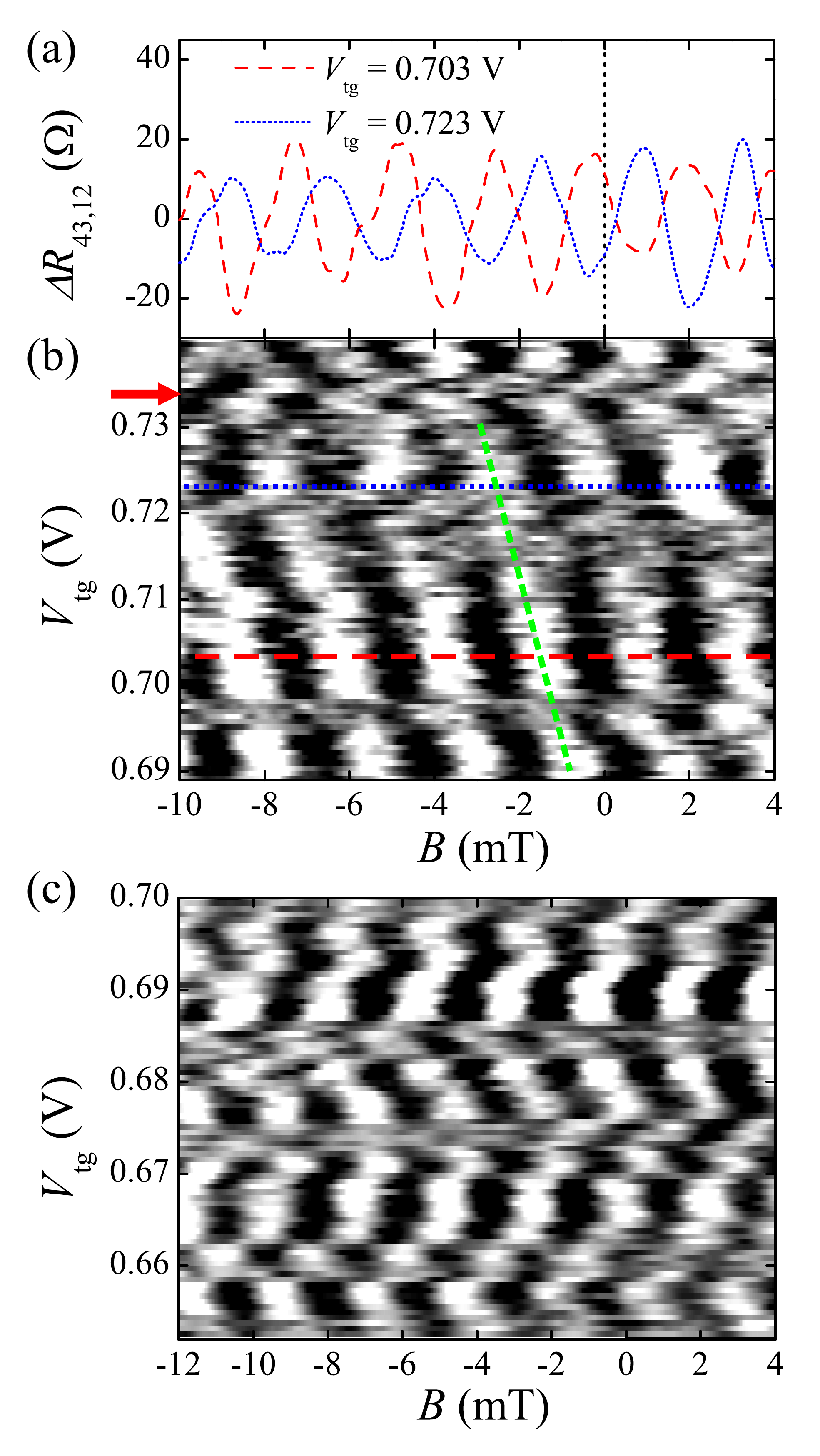}
\caption{(Color online) Oscillatory part of four-terminal magnetoresistance measurements. (a) Typical magnetoresistance for different top-gate voltages $V_{\mathrm{tg}}$ extracted from (b) as indicated by the red (dashed) and blue (dotted) lines. (b,c) Magnetoresistance in gray scale from the non-local measurement $R_{43,12}$ (b) and the local measurement $R_{41,32}$ (c) versus magnetic field and $V_{\mathrm{tg}}$. The red arrow marks a typical $\pi$-phase jump. Magnetoresistance traces were recorded for succeeding gate voltages in steps of $\Delta V_{\mathrm{tg}}=0.6$~mV at $T_\mathrm{base}=23$~mK. Ring radius $R_{\mathrm{exp}}=1$~$\mu$m.}
\label{fig:phaseshiftExp}
\end{center}
\end{figure}

\begin{figure}[t]
\begin{center}
\includegraphics[height=0.55\textheight]{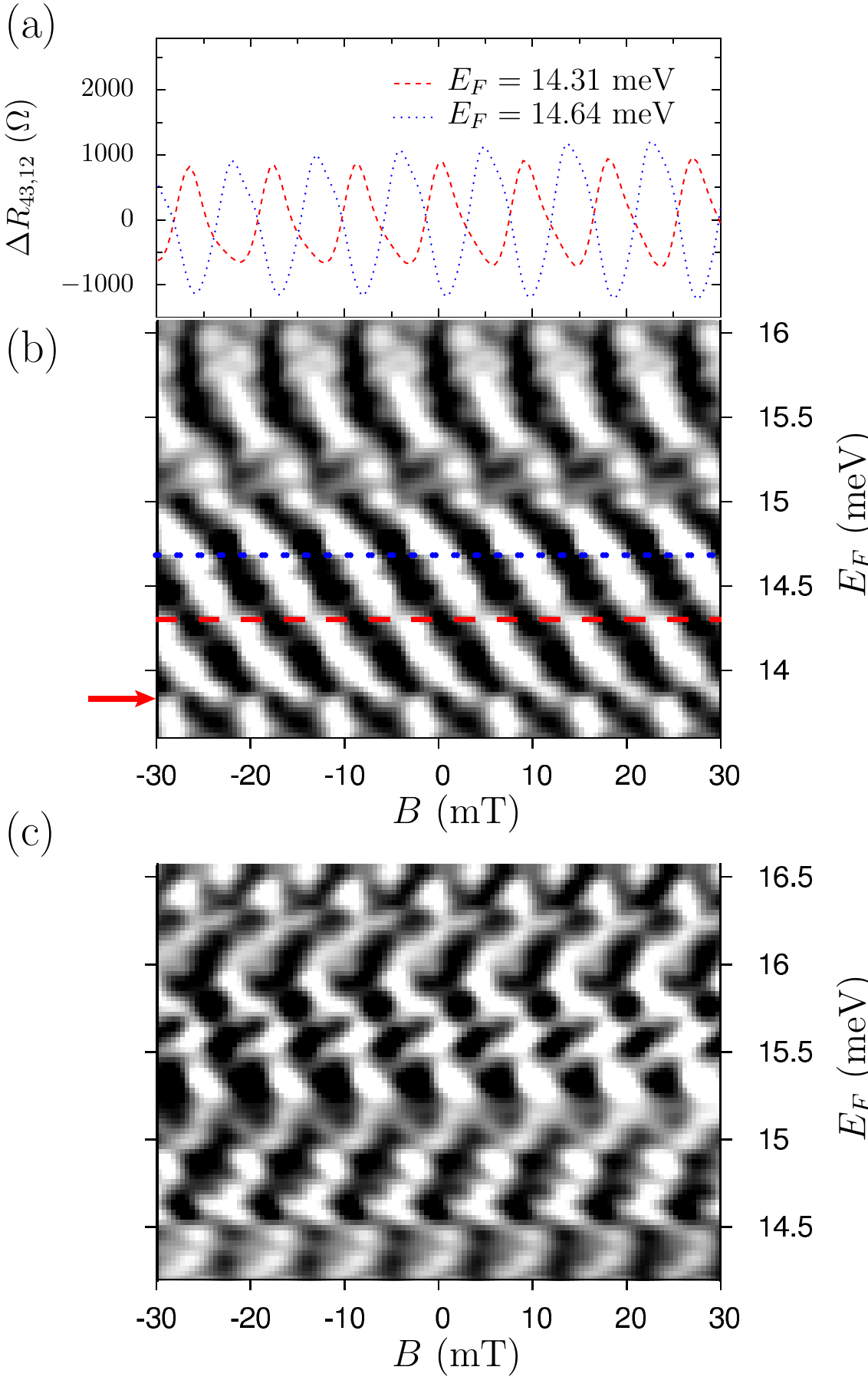}
\caption{(Color online)
Numerical simulation of ac lock-in magnetoresistance measurements. 
The effective electron temperature is set to 150~mK and the ac peak current is 6~nA.
The radii of the rounded cross-junctions are $R_{\rm left}=75$~nm for the left and
$R_{\rm right}=65$~nm for the right junction. 
(a) Oscillatory component of the non-local resistance for two different Fermi energies,
corresponding to the dashed respectively dotted lines in (b).
(b,c) Gray scale plots of the oscillatory component of the  non-local ($\Delta R_{43,12}$)  (b) and local ($\Delta R_{41,32}$) (c) resistance. The red arrow marks a phase jump of $\pi$. Ring radius $R_{\mathrm{th}}=0.5R_{\mathrm{exp}}$ (details see Sect.~\ref{sec:comparison}).
}
\label{fig:phaseshiftTheo}
\end{center}
\end{figure}

\section{Experimental Data}\label{sec:experiment}

\begin{figure}[bt]
\begin{center}
\includegraphics[width=.95\columnwidth]{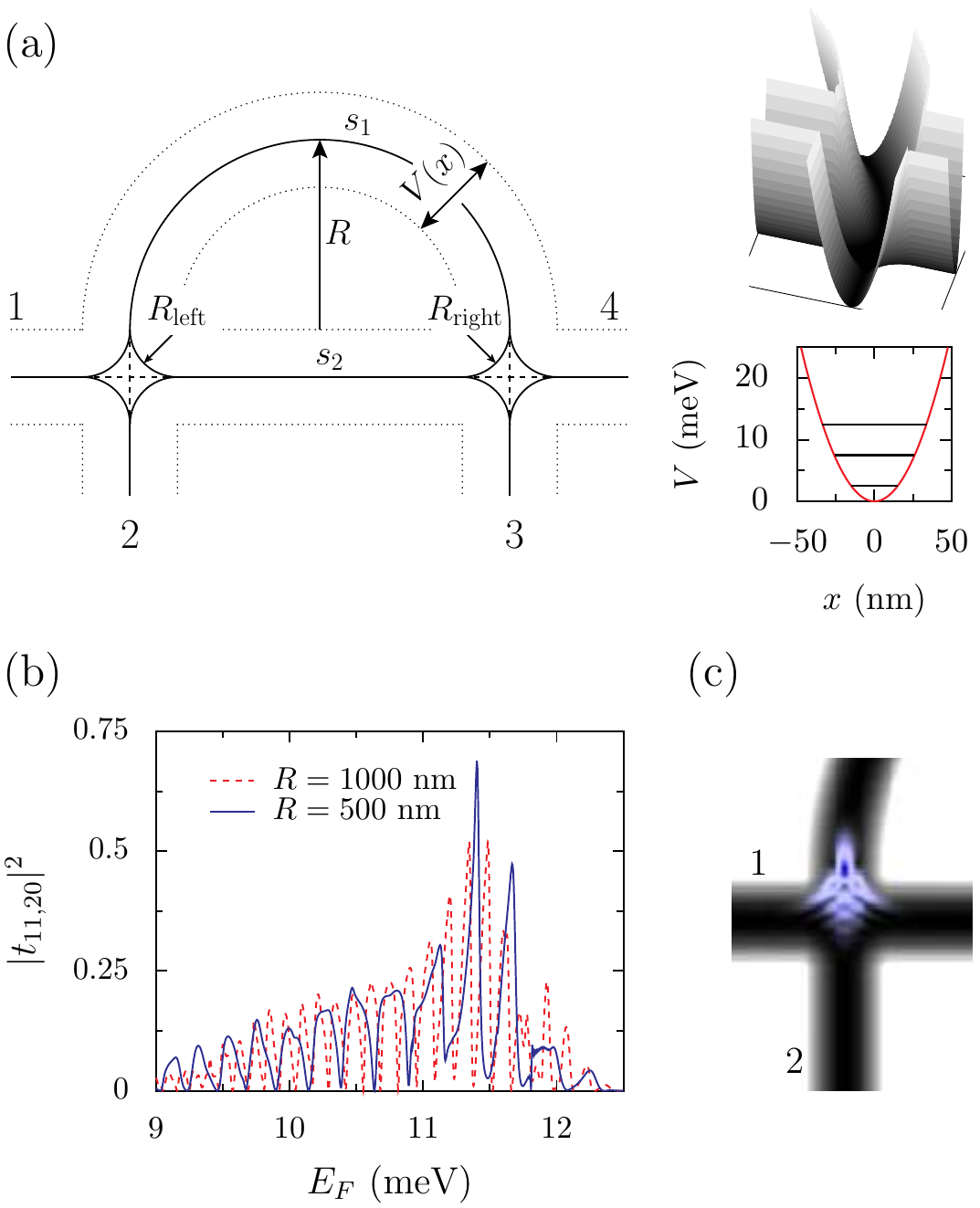}
\end{center}
\caption{\label{AharonovBohmPotential}
(Color online) (a) Geometric construction of the potential landscape. 
The AB ring of radius $R$ and the four leads are given by harmonically confined waveguides with common frequency $\omega$.
The arms $s_1$ and $s_2$ form the ring and are connected to the asymptotic leads via orthogonal cross-junctions.
Left and right cross-junction are rounded with radius $R_{\rm left}$ respectively $R_{\rm right}$.
The smooth potential of the right junction is illustrated in the 3-d plot.
(b) Fermi energy dependency of the transmission probability for inter-mode scattering from the transversal 
ground state in lead 2 to the first exited state in lead 1 ($B=0$). The level spacing of the harmonic confinement is 
set to $\hbar\omega=5$~meV, and the cross-junction radii are
$R_{\rm left}=R_{\rm right}=70$~nm. (c) Scattering at the left cross-junction of an incoming wave packet originating from lead~2 in the transverse ground state. 
}
\end{figure}

The asymmetric quantum ring is schematically depicted in Fig.~\ref{AharonovBohmPotential}~(a). We realized a waveguide geometry which allows for mode-controlled 1D transport via a global gate electrode covering the entire ring and the adjacent 2D reservoirs. A scanning electron micrograph of the device and details of the fabrication can be found in Ref.~\onlinecite{buch09}. The interferometer was designed to facilitate a comparison with theoretical calculations as shown in Figs.~\ref{fig:phaseshiftExp} (experiment) and \ref{fig:phaseshiftTheo} (theory), as well as to identify a transmission phase shift experimentally: (i) the four-terminal ring is strongly asymmetric in order to break transmission symmetries and allow for a transmission phase shift; (ii) the electron waveguides defining the ring and the leads intersect orthogonally to minimize reflections at the leads; and (iii) the 2D-1D junction connecting the 2D reservoirs and feeding the leads is located far outside the quantum ring structure and does not contribute to the four-terminal measurements.

Magnetoresistance measurements were performed with approximately 8 to 12 populated modes in small magnetic fields (up to 20~mT). Qualitatively similar results have been found for a ring with 3 to 6 populated subbands. \cite{buch10b}

The AB ring was prepared from an AlGaAs/GaAs field-effect heterostructure with a two-dimensional electron gas (2DEG) 55 nm below the surface (electron density $n_s=3.1\times10^{11}$~cm$^{-2}$, mobility $\mu=1\times10^6$~cm$^2/\mathrm{Vs}$, free mean path $l_{\mathrm{e}}\approx9.5$~$\mu$m). The geometric width of the etched waveguides amounts to 250~nm, the distances between the intersection centers of the waveguides are $s_1\approx3.3$~$\mu$m along the bent and $s_2\approx2$~$\mu$m along the straight waveguide.

We measured the four-probe resistance $R_{ij,kl}=(V_k-V_l)/I_{ij}=V_{kl}/I_{ij}$ in the local configuration, where the voltage probes are placed along the current path, e.g. $R_{41,32}$, and in the non-local configuration, where the voltage probes are separated from the current path, e.g. $R_{43,12}$. Measurements were performed with standard lock-in technique in a dilution refrigerator at the base temperature of $T_\mathrm{base}<30$~mK, which gives an effective electron temperature of approximately 100 to 150~mK. For a measurement of $R_{ij,kl}$, we fed an ac current of 12~nA rms at 73.3~Hz to terminal $i$, whereas terminal $j$ was grounded. The current was realized by a voltage of 120~mV rms from a signal generator at a resistor of 10~M$\Omega$ in series to the sample. The voltages at terminals $k$ and $l$ were measured via a preamplifier with input resistances of 100~M$\Omega$.

In order to investigate the phase sensitivity of the asymmetric quantum ring we measured the magnetoresistance as a function of the top-gate voltage $V_\mathrm{tg}$ to detect the electrostatic part of the AB-effect.
In Fig.~\ref{fig:phaseshiftExp}~(b,c) the oscillatory components of four-terminal resistance measurements are shown in gray scale versus the magnetic field and the gate voltage. Magnetoresistance measurements were recorded for successive gate voltages at $T_\mathrm{base}=23$~mK, and the background resistance was subtracted. In Fig.~\ref{fig:phaseshiftExp} (a) we depict two typical AB oscillations from Fig.~\ref{fig:phaseshiftExp} (b) at gate voltages $V_\mathrm{tg}=0.703$~V and 0.723~V. The measurements have been smoothed and the background resistance has been subtracted. The phase shift of $\Delta B/2=1.14$~mT is clearly visible and amounts to a phase of approximately $\pi$.

Fig.~\ref{fig:phaseshiftExp} (b) shows the non-local measurement $\Delta R_{43,12}$. Here, an overall resistance-oscillation phase shift is visible as indicated by the green diagonal line. The observed electrostatically induced transmission phase shift is in good agreement with a 2D estimate.\cite{buch10b} Superimposed on the overall tendency of the transmission phase are regions of reduced resolution (smaller AB amplitudes) (e.g. around $V_\mathrm{tg}=0.697$~V in Fig.~\ref{fig:phaseshiftExp} (b)), higher harmonics ($h/2e$ oscillations) and abrupt phase jumps (e.g. around $V_\mathrm{tg}=0.734$~V in Fig.~\ref{fig:phaseshiftExp} (b)). The red arrow marks the region of a typical sharp $\pi$-phase jump. 
The occurrence of a reduced amplitude, higher harmonics and abrupt phase jumps might be related to impurity scattering, electron-electron interactions or electron wave scattering and reflection in the waveguide cross-junctions. The latter possible cause would be a fundamental effect dominated by the device geometry and will be investigated in Sect.~\ref{sec:theory}.

Fig.~\ref{fig:phaseshiftExp} (c) shows the magnetoresistance gray scale plot in the local four-terminal measurement configuration, $\Delta R_{41,32}$. Here, continuous phase shifts are only occasionally visible (e.g. around $V_\mathrm{tg}=0.655$~V and 0.692~V) and their slopes in the gate voltage - magnetic field plane are different, even in sign. After a short range of gate voltage the shifts break up, and in other gate voltage ranges the phase does not change with gate voltage (e.g. around $V_\mathrm{tg}=0.665$~V). A general tendency of a phase evolution is not visible as expected in a local measurement \cite{ford90}. In contrast to non-local measurements, phase jumps occur more often and the phase seems to be restrained to evolve continuously. This is a consequence of device symmetries leading to $R_{41,32}(B)=R_{41,32}(-B)$ as is explained in the following section.

\section{Theory}\label{sec:theory}

For a realistic model of the device, we have to incorporate depletion effects along the arms of the ring and to accurately model the effectively rounded cross-junctions where scattering is strongly influenced by collimation effects.\cite{Baranger1991a} Such effects are absent in quasi-1D approaches.\cite{Gefen1984a,Buttiker1984a,Vasilopoulos2007a,Hedin2009a,Ying2008a} The eigenmode energies of the quantum wires are matched to the experimental values by using a quadratic confinement potential, which yields a constant mode separation.
The electrons in the GaAs/AlGaAs heterostructure are described within the effective mass approximation ($m^*=0.067 m_e$) and for the small magnetic fields under consideration the Zeeman splitting and spin effects are neglected. The potential profile is schematically sketched in Fig.~\ref{AharonovBohmPotential}(a). 
According to the Landauer-B\"uttiker formalism, current in lead $i$ 
\begin{equation}\label{Landauer}
 I_i=\frac{e}{h}\int_{-\infty}^{\infty}\hspace{-0.3cm}\mbox{d}E\hspace{-0.2cm} \sum_{j\neq i, n_i, n_j}\hspace{-0.3cm}|t_{in_i,jn_j}(E)|^2\,(f_i(E)-f_j(E))
\end{equation}
is related to the scattering matrix elements $t_{in_i,jn_j}$. 
The Fermi functions $f_{\gamma}=(e^{(E-\mu_{\gamma})/k_BT}+1)^{-1}$ characterize the macroscopic contacts.

The numerical effort lies in the calculation of the scattering matrix where
two major difficulties arise. Firstly, we have to compute the scattering matrix for a smooth potential 
with non-trivial topology and secondly, we need $t_{in_i,jn_j}(E)$ not only for different magnetic fields but also for a large Fermi-energy range to study the influence of the top-gate voltage. 
Several recently developed recursive Green's function methods principally allow one to compute the transmission through AB rings\cite{Kazymyrenko2008a,Wurm2010a} but yield the transmission matrices only for a single Fermi energy.
Time-dependent methods based on wave-packet dynamics have been implemented for ring structures,\cite{Chaves2009a, Szafran2005a} but suffer the disadvantage that merely the transmission of a certain pulse is detected. 
Here, we follow another approach, which is based on the combination of wave-packet methods with a Fourier analysis of the time-dependent correlation of the overlap of the wave-packets.\cite{Kramer2008a, Kramer2010b} The main advantage is that a single wave-packet run gives the energy resolved scattering-matrix elements for a large energy range, which makes this approach very efficient and well-suited for the problem at hand.

In Fig.~\ref{AharonovBohmPotential}(b) we illustrate inter-mode scattering from the transversal ground state in lead 2 to the first excited state in lead 1.
The transmission probability is a strongly fluctuating function with Fermi energy. The envelope is determined by the scattering properties of
the cross-junction (Fig.~\ref{AharonovBohmPotential}(c) shows  the scattering of a wave packet which populated the transverse ground state of lead~2 far away from the scattering region), whereas the fast varying part originates from resonances in the arms of the AB ring. 
Since electron waves can be scattered repeatedly back and forth between the two cross-junctions, the system behaves
like a Fabry-Perot interferometer and gives rise to an oscillating transmission probability. The resonance condition of maximal transmission in a Fabry-Perot interferometer is given if the length of the arm $s$ is a multiple of the half of the wavelength
\begin{equation}
  E_F-E_n =\frac{\hbar^2}{2m} \left(\frac{\pi}{s}\right)^2 i^2,
\end{equation}
where $E_F$ denotes the Fermi energy,
$E_n$ is the transversal energy to populate mode $n$ and $i$ is an integer.
Hence the energy scale of the fluctuations depends on the geometry of the AB ring and gets smaller with increasing radius $R$, see
Fig.~\ref{AharonovBohmPotential}(b). 

\begin{figure}[t]
\begin{center}
\includegraphics[width=0.95\columnwidth]{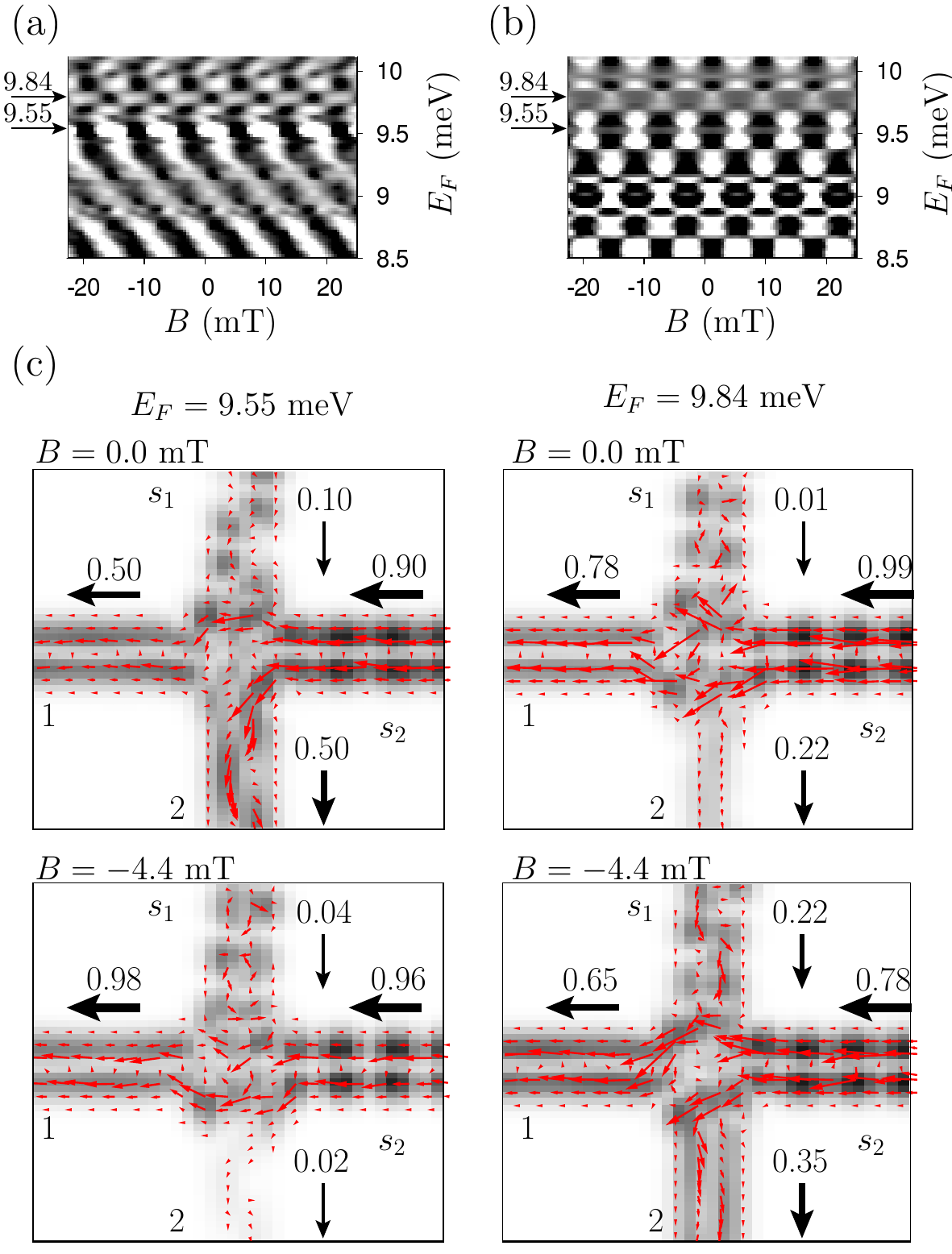}
\caption{(Color online) 
Gray scale plots of non-local ($R_{43,12}$) (a) and local ($R_{41, 32}$) (b) resistance evaluated in the linear regime. The constant background has been subtracted. 
(c) Propability density of scattering eigenstates for $E_F=9.55$~meV respectively $E_F=$9.84~meV.
The small arrows visualize the corresponding flux densities, whereas the larger arrows 
indicate the integrated flux along the transverse direction of the quantum waveguides. The flux is normalized to the
total incoming flux. The scattering behavior at the right crossing is detailed in the upper two panels for zero magnetic field and in the lower ones for $B=4.4$~mT$\approx\Delta B/2$.
$\Delta B=8.75$~mT denotes the $h/e$ AB frequency.
}
\label{EigFuFlux}
\end{center}
\end{figure}
The sensitive dependency of the transmission probabilities on the Fermi energy also leaves its mark on the AB oscillations, which show
a rich structure (see Fig.~\ref{EigFuFlux}).
Non-local and local resistance were evaluated in the linear regime $R_{mn,kl}=h/e^2(T_{km}T_{ln}-T_{kn}T_{lm})/D$ with $T_{ij}=\sum_{n_i, n_j}|t_{in_i,jn_j}(E)|^2$ and constant
$D$.\cite{Buttiker1986a}
For the considered energy range two modes are populated.
Both measurement setups show a completely different behavior. We obtain strict phase rigidity
in the local setup, where the phase of the AB oscillations
is locked either to 0 or $\pi$ at zero magnetic field. Transitions between these two values occur in form of several sharp phase jumps.
The local resistance is an even function in the magnetic field, which is a direct consequence of underlying symmetries. The scattering potential itself is mirror symmetric, leading to $T_{34}(B)=T_{21}(-B)$ and $T_{31}(B)=T_{24}(-B)$.
This symmetry is preserved by the special arrangement of voltage and current probes in the local measurement and leads in
combination with time reversal symmetry to $R_{41,32}(B)=R_{41,32}(-B)$.

The voltage probes in the non-local configuration are not arranged mirror-symmetric and thus the symmetry argument given above does not apply.
The overall tendency of the phase follows from a simplified 1D interference model. Due to the
different path lengths $s_1$ and $s_2$ the AB oscillations gain an additional phase $\Delta\alpha= k_F(s_2-s_1)$,
which depends on the longitudinal momentum $k$ and therefore on the longitudinal energy $E_F-E_n$.
If there is more than one open mode, AB oscillations
with different phases superpose each other, which can lead to abrupt phase changes as observed in the numerical
simulations.
For the present geometry ($R=500$~nm) and under the assumption that all modes contribute with equal and
energy independent weights there should be exactly one phase jump at $E_F\approx10.0$~meV within the considered energy range
of Fig.~\ref{EigFuFlux}(a). However, the sequence of jumps around $E_F=9.8$~meV is not contained in the 1D picture and require to consider inter-mode scattering.

Resonances in the arms of the ring lead to
fluctuations in the transmission probabilities, which induce fluctuations in the amplitudes of the non-local AB oscillations with Fermi energy.
Additionally, these resonances affect the scattering behavior in a more drastic way resulting in phase jumps.
This is illustrated in Fig.~\ref{EigFuFlux}(c) where we plot the probability density of scattering eigenstates
in the proximity of the left cross-junction for  energies $E_F=9.55$~meV and
$E_F=9.84$~meV, which enclose a phase jump in the non-local setup around $9.75$~meV.
In the time-dependent picture, the shown scattering eigenstates correspond to incoming electron waves 
in lead~4 populating the first exited mode. Scattering at the right cross-junction splits the waves into parts traveling along path $s_1$ respectively path
$s_2$. Both parts interfere at the left cross-junction. 
The red arrows indicate the flux density and the big black arrows illustrate the integrated flux along the transversal 
direction of the waveguides. The latter is normalized to the total incoming flux.
For zero magnetic field and $E_F=9.55$~meV, the right cross-junction distributes the incoming flux
equally to lead 1 and lead 2, whereas for $E_F=9.84$~meV transport to lead 1 dominates. 
If we increase the magnetic field to $B=4.4$~mT, which corresponds approximately to half of the $h/e$ period ($\Delta B=8.75$~mT) we find the reversed situation. Now transport to lead 2 is blocked for $E_F=9.55$~meV but
enhanced for $E_F=9.84$~meV. 
We obtain a phase shift of $\pi$ in the magnetic field dependency between these two Fermi energies. The opening and blocking of
transport is less prominent for $E_F=9.84$~meV and the AB amplitude is reduced compared to $E_F=9.55$~meV.
Note that phase jumps originating from this effect occur on the same energy scale as the resonances and can therefore
appear in sequences, see Fig.~\ref{EigFuFlux}(a) ($E_F\approx9.8$~meV). 

\section{Comparison of theoretical calculations and experimental data }\label{sec:comparison}

Scattering at the cross-junctions and thus the AB oscillation depends strongly on the transversal profile of the incoming electron waves. Whenever several modes contribute to the transport, single-mode effects superpose each other
resulting in an average behavior. 
This is confirmed experimentally where measurements with 3 to 6 open modes\cite{buch10b} are qualitatively similar to experiments with 8 to 12 populated modes (see Fig.~\ref{fig:phaseshiftExp}).
Hence single-mode effects are already averaged out with three open modes and simulations in this range are sufficient to reproduce the experimental observations of Sect.~\ref{sec:experiment}.
To reduce the numerical effort, we set the radius of the AB ring to 500~nm, which corresponds to approximately half of the experimental size.
The reduced size influences the interference in two ways. Firstly, the period of the AB oscillation increases to $\Delta B=8.75$~mT compared to
$\Delta B\approx2.5$~mT observed in the experiment and secondly, the energy scale of resonances in the arms of the ring changes, and fluctuations in the transmission amplitude occur on a larger energy scale, see
Fig.~\ref{AharonovBohmPotential}(b).
However, fundamental observations, like overall phase behavior with variation of the
Fermi energy and the occurrence of phase jumps, are not affected by halving the device size.

The open transport window of transmission amplitudes, which contribute to the current in the Landauer formula Eq.~(\ref{Landauer})
is determined by the difference of Fermi functions and thus depends on temperature and applied voltages.
The experimental currents are of the order of 12~nA leading to bias voltages of 0.03 to 0.1~mV. The effective electron
temperature was estimated to be between 100 and 150~mK, which gives rise to a thermal broadening of $4 k_BT\approx 0.05$~meV.
Since both energy scales are comparable with fluctuations in the transmission probabilities, the linear regime is not applicable and we solve the system of nonlinear equations. Nevertheless the applied currents are small enough so that 
the Landauer formula is still a good description. 

With increasing temperature and current the amplitude of the AB oscillations decreases and finer structures in the Fermi-energy dependency
smear out. Remarkably, finite currents qualitatively change the phase behavior of the AB oscillations, which becomes
especially visible in the local regime where phase rigidity gets slightly lifted. For certain energy ranges the phase
evolves continuously to lower or higher magnetic fields.
The tendency depends on the direction of the applied current and 
hence the experimental ac lock-in technique, where the measured signal is an average over negative as well as
positive currents, is taken into account in our simulations.
Additionally, we find that also details of the experimental measurement setup influence the AB oscillations. 
A symmetric voltage drop between the two current probes (push-pull configuration) leads for example to different
results than the situation where
one contact is biased and the other remains at a fixed Fermi energy.

\begin{figure}[t]
\begin{center}
\includegraphics[width=0.95\columnwidth]{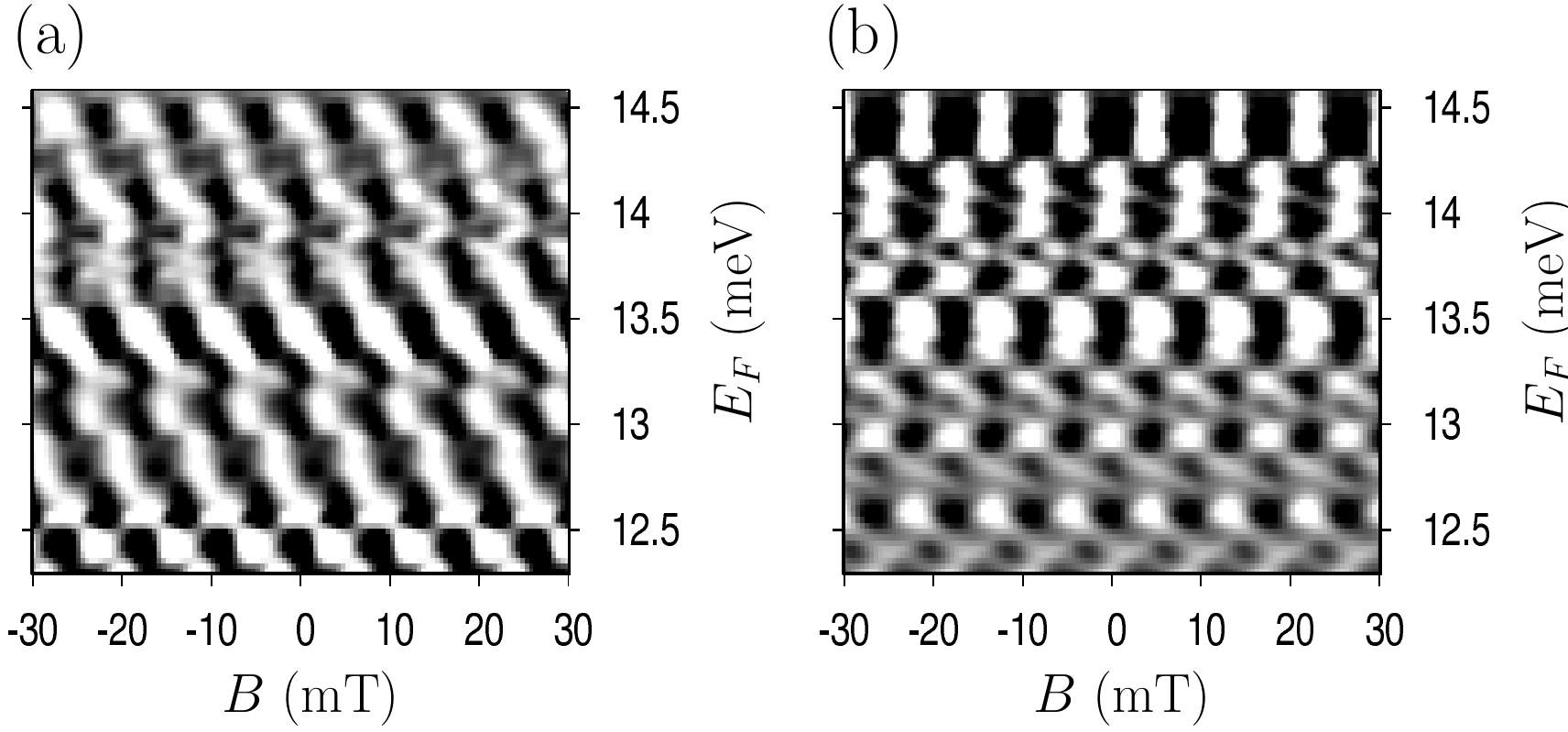}
\caption{Gray scale plots of the oscillatory components of the magnetoresistance obtained from simulations
of ac lock-in measurements. We set the effective electron temperature to 150~mK and the ac-peak current to 6~nA.
(a-b) Non-local ($R_{43,12}$) (a) and local ($R_{41,32}$) (b) resistance for a symmetric device geometry with $R_{\rm left}=R_{\rm right}=70$~ nm.
}
\label{Grayscale}
\end{center}
\end{figure}
The resistance $R_{ij,kl}$ is evaluated by applying an ac current $I_i(t)=I_{\rm max}\cos\omega t$ and measuring
the voltage $V_{kl}(t)=(\mu_k(t)-\mu_l(t))/e$. 
The contacts $k$ and $l$ are perfect voltage probes, forcing the currents 
$I_k$ and $I_l$ to vanish. The chemical potential of the contact $j$ is fixed to $\mu_j=E_F$.
We divide the ac oscillation period into discrete time steps. For each step we solve 
a nonlinear system of equations, whose solution gives the chemical potentials of all contacts.  
The ac lock-in amplifier detects the integrated (rms) signal
\begin{equation}
 V_{kl,\mbox{\scriptsize rms}}=\frac{\omega}{\sqrt{2}\,\pi}\int_0^{2\pi/\omega}V_{kl}(I_{ij}(t))\cos\omega t\ \mbox{d}t
\end{equation}
which determines the resistance $R_{ij,kl}=V_{kl,{\rm rms}}/I_{\rm rms}$ ($I_{\rm rms}=I_{\rm max}/\sqrt{2}$).

In Fig.~\ref{Grayscale} we show the numerical results for the oscillatory components of the non-local and local resistance versus Fermi energy with the constant background subtracted. 
Based on Shubnikov-de Haas-measurements of the electron density in the ring,\cite{buch10b} we estimate that the experimental data illustrated in Fig.~\ref{fig:phaseshiftExp} 
cover an energy range of 1.6~meV which is comparable to the simulated range of 2.5~meV.

The non-local setup, Fig.~\ref{Grayscale}(a), shows a continuous phase shift towards higher magnetic fields. This general tendency is interrupted by
phase jumps of $\pi$, for example at 12.5~meV. 
Compared to the linear regime, finer structures in the gray scale plot are thermally smeared out.
The situation for the local setup, illustrated in Fig.~\ref{Grayscale}(b) is different. Here the symmetry of the device results in phase rigidity.
However, in contrast to the linear regime (Fig.~\ref{EigFuFlux}(b)), strict phase rigidity is slightly lifted because of the applied finite current.
For example, we obtain a continuous phase change in a small region around 13.2~meV. 
Areas of reduced resolution alternate with regions of higher AB amplitudes. This is a consequence of the energy
dependency of the scattering behavior of the orthogonal cross-junctions.  The phase jumps, which appear on a 
faster energy scale are due to resonances in the arms of the ring, as discussed above. 

So far the general phase behavior and the presence of phase jumps are reproduced from numerical simulations.
The remaining question concerns the experimentally observed continuous phase shift for certain gate-voltage ranges in the local setup which
cannot be explained by the presence of a finite current alone. Note that Fig.~\ref{Grayscale}(b) still shows large regions 
where the phase is locked.
Phase rigidity is a fundamental property incorporated in the symmetry of the device. If this symmetry is broken, phase rigidity is not expected to be present anymore.
In the experimental setup there are several possible sources which can break symmetry, such as disorder, impurities or asymmetries due to fabrication processes of the device. 

In the following we assume a small asymmetry
between the two orthogonal cross-junctions and we choose different radii for the rounded junctions.
The radius for the left junction was set to $R_{\rm left}=75$~nm, whereas we used $R_{\rm right}=65$~nm for the right crossing. 
The corresponding results for the non-local and local setup are illustrated in Fig.~\ref{fig:phaseshiftTheo}. 
The phase of the non-local measurement (Fig.~\ref{fig:phaseshiftTheo}(b)) shows qualitatively the same behavior as the one of the symmetric device ($R_{\rm left}=R_{\rm right}$). It
shifts continuously and is interrupted by phase jumps, e.g. at $E_F\approx13.9$~meV as indicated by the red arrow. Fig.~\ref{fig:phaseshiftTheo}(a) shows the oscillatory components for the non-local resistance for two different Fermi
energies, which correspond to the red (dashed) and blue (dotted) lines in the gray scale plot. Between $E_F=14.31$~meV and $E_F=14.64$~meV
the phase of the AB oscillation undergoes a continuous shift of $\pi$. 

For the local setup, the phase evolution for the asymmetric case is
shown in Fig. 2(c). Comparing Fig. 4 (b) for the symmetric junctions in
the linear regime, Fig. 5 (b) (symmetric junctions at finite bias), and
Fig. 2 (c), we find that both effects (finite bias and device
asymmetries) have to be included in our model to obtain regions of phase
drifts, visible in the experiment (Fig. 1 (c)). In contrast to the
non-local setup, no preferred direction of the phase shift exists.

The simulated AB amplitudes (Fig.~\ref{fig:phaseshiftTheo}) are larger than the ones in Fig.~\ref{fig:phaseshiftExp} and the experimentally measured
amplitude is well below the theoretical prediction. There are several mechanisms which can explain the experimentally reduced AB oscillations.
Besides thermal averaging, coupling to a thermal environment (electron-phonon interaction)
gives rise to decoherence and is another source of temperature induced dephasing which was not included in the simulations.
Recent theoretical investigations\cite{Seelig2001a,Levkivskyi2008a,Kovrizhin2009a,Kotimaki2010a} also propose that electron-electron interactions reduce the measured AB signals.
A more detailed discussion can be found in Ref.~\onlinecite{buch10b}.

\section{Summary}

We have investigated the electron transmission phase and the origin of irregular phase jumps in an four-terminal AB interferometer in combined experimental and theoretical approaches. Our waveguide ring design allows for a transmission phase shift via the Fermi wave number. The realistic theoretical model takes into account experimental conditions as the confinement potential and the ac lock-in measurement circuitry. Both experimental and theoretical results show that phase rigidity remains largely intact in the local measurement configuration but is interrupted in some regions with continuous phase shifts due to the ac measurement technique and asymmetries in the cross-junctions arising from small imperfections in the fabrication process. In contrast, the phase evolves continuously in the non-local measurement due to the broken symmetry. Irregular phase jumps by $\pi$ occur in both measurement configurations in the experimental as well as the theoretical approach. The investigation of Fermi-energy-dependent scattering probabilities reveals that $\pi$-jumps are caused by the strong scattering resonances within the junctions which redirect the current flow and lead to multiple reflections in the arms of the ring. Consequently, an AB-interferometer based phase detector requires minimized scattering and a symmetry-breaking measurement setup. In our waveguide AB ring scattering is reduced by the implementation of orthogonal waveguide cross-junctions.

We find that single-mode effects are mostly washed out when several (8-12) modes are populated. Thus a further investigation calls for single-mode transport to get an insight into single-mode interference properties in electron waveguide ring structures.

\section*{Acknowledgments}

SSB and SSF gratefully acknowledge financial support from the Deutsche Forschungsgemeinschaft within the priority program SPP1285. SSB appreciates support by the Research School of the Ruhr-Universi\"at Bochum. SFF gratefully acknowledges support by the Alexander-von-Humboldt Foundation. TK and CK appreciate funding by the Emmy-Noether program of the DFG, grant KR 2889/2 and helpful discussions with V.~Kr\"uckl and K.~Richter.


\begin{thebibliography}{41}
\expandafter\ifx\csname natexlab\endcsname\relax\def\natexlab#1{#1}\fi
\expandafter\ifx\csname bibnamefont\endcsname\relax
  \def\bibnamefont#1{#1}\fi
\expandafter\ifx\csname bibfnamefont\endcsname\relax
  \def\bibfnamefont#1{#1}\fi
\expandafter\ifx\csname citenamefont\endcsname\relax
  \def\citenamefont#1{#1}\fi
\expandafter\ifx\csname url\endcsname\relax
  \def\url#1{\texttt{#1}}\fi
\expandafter\ifx\csname urlprefix\endcsname\relax\def\urlprefix{URL }\fi
\providecommand{\bibinfo}[2]{#2}
\providecommand{\eprint}[2][]{\url{#2}}

\bibitem[{\citenamefont{Schuster et~al.}(1997)\citenamefont{Schuster, Buks,
  Heiblum, Mahalu, Umansky, and Shtrikman}}]{schu97}
\bibinfo{author}{\bibfnamefont{R.}~\bibnamefont{Schuster}},
  \bibinfo{author}{\bibfnamefont{E.}~\bibnamefont{Buks}},
  \bibinfo{author}{\bibfnamefont{M.}~\bibnamefont{Heiblum}},
  \bibinfo{author}{\bibfnamefont{D.}~\bibnamefont{Mahalu}},
  \bibinfo{author}{\bibfnamefont{V.}~\bibnamefont{Umansky}}, \bibnamefont{and}
  \bibinfo{author}{\bibfnamefont{H.}~\bibnamefont{Shtrikman}},
  \bibinfo{journal}{Nature} \textbf{\bibinfo{volume}{385}},
  \bibinfo{pages}{417} (\bibinfo{year}{1997}).

\bibitem[{\citenamefont{Katsumoto}(2007)}]{kats07}
\bibinfo{author}{\bibfnamefont{S.}~\bibnamefont{Katsumoto}},
  \bibinfo{journal}{Journal of Physics: Condensed Matter}
  \textbf{\bibinfo{volume}{19}}, \bibinfo{pages}{233201}
  (\bibinfo{year}{2007}).

\bibitem[{\citenamefont{Yacoby et~al.}(1995)\citenamefont{Yacoby, Heiblum,
  Mahalu, and Shtrikman}}]{yac95}
\bibinfo{author}{\bibfnamefont{A.}~\bibnamefont{Yacoby}},
  \bibinfo{author}{\bibfnamefont{M.}~\bibnamefont{Heiblum}},
  \bibinfo{author}{\bibfnamefont{D.}~\bibnamefont{Mahalu}}, \bibnamefont{and}
  \bibinfo{author}{\bibfnamefont{H.}~\bibnamefont{Shtrikman}},
  \bibinfo{journal}{Phys. Rev. Lett.} \textbf{\bibinfo{volume}{74}},
  \bibinfo{pages}{4047} (\bibinfo{year}{1995}).

\bibitem[{\citenamefont{Kobayashi
  et~al.}(2002{\natexlab{a}})\citenamefont{Kobayashi, Aikawa, Katsumoto, and
  Iye}}]{koba0204a}
\bibinfo{author}{\bibfnamefont{K.}~\bibnamefont{Kobayashi}},
  \bibinfo{author}{\bibfnamefont{H.}~\bibnamefont{Aikawa}},
  \bibinfo{author}{\bibfnamefont{S.}~\bibnamefont{Katsumoto}},
  \bibnamefont{and} \bibinfo{author}{\bibfnamefont{Y.}~\bibnamefont{Iye}},
  \bibinfo{journal}{Phys. Rev. Lett.} \textbf{\bibinfo{volume}{88}},
  \bibinfo{pages}{256806} (\bibinfo{year}{2002}{\natexlab{a}}).

\bibitem[{\citenamefont{Kobayashi et~al.}(2004)\citenamefont{Kobayashi, Aikawa,
  Sano, Katsumoto, and Iye}}]{koba0204b}
\bibinfo{author}{\bibfnamefont{K.}~\bibnamefont{Kobayashi}},
  \bibinfo{author}{\bibfnamefont{H.}~\bibnamefont{Aikawa}},
  \bibinfo{author}{\bibfnamefont{A.}~\bibnamefont{Sano}},
  \bibinfo{author}{\bibfnamefont{S.}~\bibnamefont{Katsumoto}},
  \bibnamefont{and} \bibinfo{author}{\bibfnamefont{Y.}~\bibnamefont{Iye}},
  \bibinfo{journal}{Phys. Rev. B} \textbf{\bibinfo{volume}{70}},
  \bibinfo{pages}{035319} (\bibinfo{year}{2004}).

\bibitem[{\citenamefont{Sigrist et~al.}(2004)\citenamefont{Sigrist, Fuhrer,
  Ihn, Ensslin, Ulloa, Wegscheider, and Bichler}}]{sigr04}
\bibinfo{author}{\bibfnamefont{M.}~\bibnamefont{Sigrist}},
  \bibinfo{author}{\bibfnamefont{A.}~\bibnamefont{Fuhrer}},
  \bibinfo{author}{\bibfnamefont{T.}~\bibnamefont{Ihn}},
  \bibinfo{author}{\bibfnamefont{K.}~\bibnamefont{Ensslin}},
  \bibinfo{author}{\bibfnamefont{S.~E.} \bibnamefont{Ulloa}},
  \bibinfo{author}{\bibfnamefont{W.}~\bibnamefont{Wegscheider}},
  \bibnamefont{and} \bibinfo{author}{\bibfnamefont{M.}~\bibnamefont{Bichler}},
  \bibinfo{journal}{Phys. Rev. Lett.} \textbf{\bibinfo{volume}{93}},
  \bibinfo{pages}{066802} (\bibinfo{year}{2004}).

\bibitem[{\citenamefont{Y.~Ji et~al.}(2000)\citenamefont{Y.~Ji, Sprinzak,
  Mahalu, and Shtrikman}}]{ji00a}
\bibinfo{author}{\bibfnamefont{M.~H.} \bibnamefont{Y.~Ji}},
  \bibinfo{author}{\bibfnamefont{D.}~\bibnamefont{Sprinzak}},
  \bibinfo{author}{\bibfnamefont{D.}~\bibnamefont{Mahalu}}, \bibnamefont{and}
  \bibinfo{author}{\bibfnamefont{H.}~\bibnamefont{Shtrikman}},
  \bibinfo{journal}{Science} \textbf{\bibinfo{volume}{290}},
  \bibinfo{pages}{779} (\bibinfo{year}{2000}).

\bibitem[{\citenamefont{Ji et~al.}(2002)\citenamefont{Ji, Heiblum, and
  Shtrikman}}]{ji00b}
\bibinfo{author}{\bibfnamefont{Y.}~\bibnamefont{Ji}},
  \bibinfo{author}{\bibfnamefont{M.}~\bibnamefont{Heiblum}}, \bibnamefont{and}
  \bibinfo{author}{\bibfnamefont{H.}~\bibnamefont{Shtrikman}},
  \bibinfo{journal}{Phys. Rev. Lett.} \textbf{\bibinfo{volume}{88}},
  \bibinfo{pages}{076601} (\bibinfo{year}{2002}).

\bibitem[{\citenamefont{Avinun-Kalish et~al.}(2005)\citenamefont{Avinun-Kalish,
  Heiblum, Zarchin, Mahalu, and Umansky}}]{ji00c}
\bibinfo{author}{\bibfnamefont{M.}~\bibnamefont{Avinun-Kalish}},
  \bibinfo{author}{\bibfnamefont{M.}~\bibnamefont{Heiblum}},
  \bibinfo{author}{\bibfnamefont{O.}~\bibnamefont{Zarchin}},
  \bibinfo{author}{\bibfnamefont{D.}~\bibnamefont{Mahalu}}, \bibnamefont{and}
  \bibinfo{author}{\bibfnamefont{V.}~\bibnamefont{Umansky}},
  \bibinfo{journal}{Nature} \textbf{\bibinfo{volume}{436}},
  \bibinfo{pages}{529} (\bibinfo{year}{2005}).

\bibitem[{\citenamefont{Zaffalon et~al.}(2008)\citenamefont{Zaffalon, Bid,
  Heiblum, Mahalu, and Umansky}}]{ji00d}
\bibinfo{author}{\bibfnamefont{M.}~\bibnamefont{Zaffalon}},
  \bibinfo{author}{\bibfnamefont{A.}~\bibnamefont{Bid}},
  \bibinfo{author}{\bibfnamefont{M.}~\bibnamefont{Heiblum}},
  \bibinfo{author}{\bibfnamefont{D.}~\bibnamefont{Mahalu}}, \bibnamefont{and}
  \bibinfo{author}{\bibfnamefont{V.}~\bibnamefont{Umansky}},
  \bibinfo{journal}{Phys. Rev. Lett.} \textbf{\bibinfo{volume}{100}},
  \bibinfo{pages}{226601} (\bibinfo{year}{2008}).

\bibitem[{\citenamefont{Oreg and Gefen}(1997)}]{Oreg1997a}
\bibinfo{author}{\bibfnamefont{Y.}~\bibnamefont{Oreg}} \bibnamefont{and}
  \bibinfo{author}{\bibfnamefont{Y.}~\bibnamefont{Gefen}},
  \bibinfo{journal}{Phys. Rev. B} \textbf{\bibinfo{volume}{55}},
  \bibinfo{pages}{13726} (\bibinfo{year}{1997}).

\bibitem[{\citenamefont{Entin-Wohlman et~al.}(2002)\citenamefont{Entin-Wohlman,
  Aharony, Imry, Levinson, and Schiller}}]{enti02}
\bibinfo{author}{\bibfnamefont{O.}~\bibnamefont{Entin-Wohlman}},
  \bibinfo{author}{\bibfnamefont{A.}~\bibnamefont{Aharony}},
  \bibinfo{author}{\bibfnamefont{Y.}~\bibnamefont{Imry}},
  \bibinfo{author}{\bibfnamefont{Y.}~\bibnamefont{Levinson}}, \bibnamefont{and}
  \bibinfo{author}{\bibfnamefont{A.}~\bibnamefont{Schiller}},
  \bibinfo{journal}{Phys. Rev. Lett.} \textbf{\bibinfo{volume}{88}},
  \bibinfo{pages}{166801} (\bibinfo{year}{2002}).

\bibitem[{\citenamefont{Silva et~al.}(2002)\citenamefont{Silva, Oreg, and
  Gefen}}]{Silva2002a}
\bibinfo{author}{\bibfnamefont{A.}~\bibnamefont{Silva}},
  \bibinfo{author}{\bibfnamefont{Y.}~\bibnamefont{Oreg}}, \bibnamefont{and}
  \bibinfo{author}{\bibfnamefont{Y.}~\bibnamefont{Gefen}},
  \bibinfo{journal}{Phys. Rev. B} \textbf{\bibinfo{volume}{66}},
  \bibinfo{pages}{195316} (\bibinfo{year}{2002}).

\bibitem[{\citenamefont{Aharonov and Bohm}(1959)}]{ahar59}
\bibinfo{author}{\bibfnamefont{Y.}~\bibnamefont{Aharonov}} \bibnamefont{and}
  \bibinfo{author}{\bibfnamefont{D.}~\bibnamefont{Bohm}},
  \bibinfo{journal}{Phys. Rev.} \textbf{\bibinfo{volume}{115}},
  \bibinfo{pages}{485} (\bibinfo{year}{1959}).

\bibitem[{\citenamefont{Webb et~al.}(1985)\citenamefont{Webb, Washburn, Umbach,
  and Laibowitz}}]{webb85}
\bibinfo{author}{\bibfnamefont{R.~A.} \bibnamefont{Webb}},
  \bibinfo{author}{\bibfnamefont{S.}~\bibnamefont{Washburn}},
  \bibinfo{author}{\bibfnamefont{C.~P.} \bibnamefont{Umbach}},
  \bibnamefont{and} \bibinfo{author}{\bibfnamefont{R.~B.}
  \bibnamefont{Laibowitz}}, \bibinfo{journal}{Phys. Rev. Lett.}
  \textbf{\bibinfo{volume}{54}}, \bibinfo{pages}{2696} (\bibinfo{year}{1985}).

\bibitem[{\citenamefont{Kobayashi
  et~al.}(2002{\natexlab{b}})\citenamefont{Kobayashi, Aikawa, Katsumoto, and
  Iye}}]{koba02}
\bibinfo{author}{\bibfnamefont{K.}~\bibnamefont{Kobayashi}},
  \bibinfo{author}{\bibfnamefont{H.}~\bibnamefont{Aikawa}},
  \bibinfo{author}{\bibfnamefont{S.}~\bibnamefont{Katsumoto}},
  \bibnamefont{and} \bibinfo{author}{\bibfnamefont{Y.}~\bibnamefont{Iye}},
  \bibinfo{journal}{J. Phys. Soc. Jpn.} \textbf{\bibinfo{volume}{71}},
  \bibinfo{pages}{2094} (\bibinfo{year}{2002}{\natexlab{b}}).

\bibitem[{\citenamefont{Onsager}(1931)}]{onsa31}
\bibinfo{author}{\bibfnamefont{L.}~\bibnamefont{Onsager}},
  \bibinfo{journal}{Phys. Rev.} \textbf{\bibinfo{volume}{38}},
  \bibinfo{pages}{2265} (\bibinfo{year}{1931}).

\bibitem[{\citenamefont{Casimir}(1945)}]{casi45}
\bibinfo{author}{\bibfnamefont{H.~B.~G.} \bibnamefont{Casimir}},
  \bibinfo{journal}{Rev. Mod. Phys.} \textbf{\bibinfo{volume}{17}},
  \bibinfo{pages}{343} (\bibinfo{year}{1945}).

\bibitem[{\citenamefont{Buttiker}(1988)}]{butt88}
\bibinfo{author}{\bibfnamefont{M.} \bibnamefont{B\"uttiker}},
  \bibinfo{journal}{IBM J. Res. Develop.} \textbf{\bibinfo{volume}{32}},
  \bibinfo{pages}{317} (\bibinfo{year}{1988}).

\bibitem[{\citenamefont{Pedersen et~al.}(1999)\citenamefont{Pedersen, Hansen,
  Kristensen, S$\o$rensen, and Lindelof}}]{Pedersen1999a}
\bibinfo{author}{\bibfnamefont{S.}~\bibnamefont{Pedersen}},
  \bibinfo{author}{\bibfnamefont{A.~E.} \bibnamefont{Hansen}},
  \bibinfo{author}{\bibfnamefont{A.}~\bibnamefont{Kristensen}},
  \bibinfo{author}{\bibfnamefont{C.~B.} \bibnamefont{S$\o$rensen}},
  \bibnamefont{and} \bibinfo{author}{\bibfnamefont{P.~E.}
  \bibnamefont{Lindelof}}, \bibinfo{journal}{Phys. Rev. B}
  \textbf{\bibinfo{volume}{61}}, \bibinfo{pages}{5457} (\bibinfo{year}{1999}).

\bibitem[{\citenamefont{Cernicchiaro et~al.}(1997)\citenamefont{Cernicchiaro,
  Martin, Hasselbach, Mailly, and Benoit}}]{Cernicchiaro1997}
\bibinfo{author}{\bibfnamefont{G.}~\bibnamefont{Cernicchiaro}},
  \bibinfo{author}{\bibfnamefont{T.}~\bibnamefont{Martin}},
  \bibinfo{author}{\bibfnamefont{K.}~\bibnamefont{Hasselbach}},
  \bibinfo{author}{\bibfnamefont{D.}~\bibnamefont{Mailly}}, \bibnamefont{and}
  \bibinfo{author}{\bibfnamefont{A.}~\bibnamefont{Benoit}},
  \bibinfo{journal}{Phys. Rev. Lett.} \textbf{\bibinfo{volume}{79}},
  \bibinfo{pages}{273} (\bibinfo{year}{1997}).

\bibitem[{\citenamefont{Gefen et~al.}(1984)\citenamefont{Gefen, Imry, and
  Azbel}}]{Gefen1984a}
\bibinfo{author}{\bibfnamefont{Y.}~\bibnamefont{Gefen}},
  \bibinfo{author}{\bibfnamefont{Y.}~\bibnamefont{Imry}}, \bibnamefont{and}
  \bibinfo{author}{\bibfnamefont{M.~Y.} \bibnamefont{Azbel}},
  \bibinfo{journal}{Phys. Rev. Lett.} \textbf{\bibinfo{volume}{52}},
  \bibinfo{pages}{129} (\bibinfo{year}{1984}).

\bibitem[{\citenamefont{B\"uttiker et~al.}(1984)\citenamefont{B\"uttiker, Imry,
  and Azbel}}]{Buttiker1984a}
\bibinfo{author}{\bibfnamefont{M.}~\bibnamefont{B\"uttiker}},
  \bibinfo{author}{\bibfnamefont{Y.}~\bibnamefont{Imry}}, \bibnamefont{and}
  \bibinfo{author}{\bibfnamefont{M.~Y.} \bibnamefont{Azbel}},
  \bibinfo{journal}{Phys. Rev. A} \textbf{\bibinfo{volume}{30}},
  \bibinfo{pages}{1982} (\bibinfo{year}{1984}).

\bibitem[{\citenamefont{Vasilopoulos et~al.}(2007)\citenamefont{Vasilopoulos,
  K\'alm\'an, Peeters, and Benedict}}]{Vasilopoulos2007a}
\bibinfo{author}{\bibfnamefont{P.}~\bibnamefont{Vasilopoulos}},
  \bibinfo{author}{\bibfnamefont{O.}~\bibnamefont{K\'alm\'an}},
  \bibinfo{author}{\bibfnamefont{F.~M.} \bibnamefont{Peeters}},
  \bibnamefont{and} \bibinfo{author}{\bibfnamefont{M.~G.}
  \bibnamefont{Benedict}}, \bibinfo{journal}{Phys. Rev. B}
  \textbf{\bibinfo{volume}{75}}, \bibinfo{pages}{035304}
  (\bibinfo{year}{2007}).

\bibitem[{\citenamefont{Hedin et~al.}(2009)\citenamefont{Hedin, Joe, and
  Satanin}}]{Hedin2009a}
\bibinfo{author}{\bibfnamefont{E.~R.} \bibnamefont{Hedin}},
  \bibinfo{author}{\bibfnamefont{Y.~S.} \bibnamefont{Joe}}, \bibnamefont{and}
  \bibinfo{author}{\bibfnamefont{A.~M.} \bibnamefont{Satanin}},
  \bibinfo{journal}{Journal of Physics Condensed Matter}
  \textbf{\bibinfo{volume}{21}}, \bibinfo{pages}{015303}
  (\bibinfo{year}{2009}).

\bibitem[{\citenamefont{Ying et~al.}(2008)\citenamefont{Ying, Jin, and
  Ma}}]{Ying2008a}
\bibinfo{author}{\bibfnamefont{Y.}~\bibnamefont{Ying}},
  \bibinfo{author}{\bibfnamefont{G.}~\bibnamefont{Jin}}, \bibnamefont{and}
  \bibinfo{author}{\bibfnamefont{Y.-Q.} \bibnamefont{Ma}},
  \bibinfo{journal}{Europhys. Lett.} \textbf{\bibinfo{volume}{84}},
  \bibinfo{pages}{67012} (\bibinfo{year}{2008}).

\bibitem[{\citenamefont{Kramer et~al.}(2010)\citenamefont{Kramer, Kreisbeck,
  and Krueckl}}]{Kramer2010b}
\bibinfo{author}{\bibfnamefont{T.}~\bibnamefont{Kramer}},
  \bibinfo{author}{\bibfnamefont{C.}~\bibnamefont{Kreisbeck}},
  \bibnamefont{and} \bibinfo{author}{\bibfnamefont{V.}~\bibnamefont{Krueckl}},
  \bibinfo{journal}{Physica Scripta, in press (arXiv} \bibinfo{pages}{1002.5042)}
  (\bibinfo{year}{2010}).

\bibitem[{\citenamefont{Kramer et~al.}(2008)\citenamefont{Kramer, Heller, and
  Parrott}}]{Kramer2008a}
\bibinfo{author}{\bibfnamefont{T.}~\bibnamefont{Kramer}},
  \bibinfo{author}{\bibfnamefont{E.~J.} \bibnamefont{Heller}},
  \bibnamefont{and} \bibinfo{author}{\bibfnamefont{R.~E.}
  \bibnamefont{Parrott}}, \bibinfo{journal}{J. Phys. Conf. Ser.}
  \textbf{\bibinfo{volume}{99}}, \bibinfo{pages}{012010}
  (\bibinfo{year}{2008}).

\bibitem[{\citenamefont{Buchholz et~al.}(2009)\citenamefont{Buchholz, Fischer,
  Kunze, Reuter, and Wieck}}]{buch09}
\bibinfo{author}{\bibfnamefont{S.~S.} \bibnamefont{Buchholz}},
  \bibinfo{author}{\bibfnamefont{S.~F.} \bibnamefont{Fischer}},
  \bibinfo{author}{\bibfnamefont{U.}~\bibnamefont{Kunze}},
  \bibinfo{author}{\bibfnamefont{D.}~\bibnamefont{Reuter}}, \bibnamefont{and}
  \bibinfo{author}{\bibfnamefont{A.~D.} \bibnamefont{Wieck}},
  \bibinfo{journal}{Applied Physics Letters} \textbf{\bibinfo{volume}{94}},
  \bibinfo{eid}{022107} (\bibinfo{year}{2009}).

\bibitem[{\citenamefont{Buchholz et~al.}(2010)\citenamefont{Buchholz, Fischer,
  Kunze, Bell, Reuter, and Wieck}}]{buch10b}
\bibinfo{author}{\bibfnamefont{S.~S.}~\bibnamefont{Buchholz}},
  \bibinfo{author}{\bibfnamefont{S.~F.}~\bibnamefont{Fischer}},
  \bibinfo{author}{\bibfnamefont{U.}~\bibnamefont{Kunze}},
  \bibinfo{author}{\bibfnamefont{M.}~\bibnamefont{Bell}},
  \bibinfo{author}{\bibfnamefont{D.}~\bibnamefont{Reuter}}, \bibnamefont{and}
  \bibinfo{author}{\bibfnamefont{A.~D.} \bibnamefont{Wieck}},
  \bibinfo{journal}{arXiv} \bibinfo{pages}{1005.2081} (\bibinfo{year}{2010}).

\bibitem[{\citenamefont{Ford et~al.}(1990)\citenamefont{Ford, Fowler, Hong,
  Knoedler, Laux, Wainer, and Washburn}}]{ford90}
\bibinfo{author}{\bibfnamefont{C.}~\bibnamefont{Ford}},
  \bibinfo{author}{\bibfnamefont{A.}~\bibnamefont{Fowler}},
  \bibinfo{author}{\bibfnamefont{J.}~\bibnamefont{Hong}},
  \bibinfo{author}{\bibfnamefont{C.}~\bibnamefont{Knoedler}},
  \bibinfo{author}{\bibfnamefont{S.}~\bibnamefont{Laux}},
  \bibinfo{author}{\bibfnamefont{J.}~\bibnamefont{Wainer}}, \bibnamefont{and}
  \bibinfo{author}{\bibfnamefont{S.}~\bibnamefont{Washburn}},
  \bibinfo{journal}{Surface Science} \textbf{\bibinfo{volume}{229}},
  \bibinfo{pages}{307 } (\bibinfo{year}{1990}).

\bibitem[{\citenamefont{Baranger et~al.}(1991)\citenamefont{Baranger,
  DiVincenzo, Jalabert, and Stone}}]{Baranger1991a}
\bibinfo{author}{\bibfnamefont{H.~U.} \bibnamefont{Baranger}},
  \bibinfo{author}{\bibfnamefont{D.~P.} \bibnamefont{DiVincenzo}},
  \bibinfo{author}{\bibfnamefont{R.~A.} \bibnamefont{Jalabert}},
  \bibnamefont{and} \bibinfo{author}{\bibfnamefont{A.~D.} \bibnamefont{Stone}},
  \bibinfo{journal}{Phys. Rev. B} \textbf{\bibinfo{volume}{44}},
  \bibinfo{pages}{10637} (\bibinfo{year}{1991}).

\bibitem[{\citenamefont{Kazymyrenko and Waintal}(2008)}]{Kazymyrenko2008a}
\bibinfo{author}{\bibfnamefont{K.}~\bibnamefont{Kazymyrenko}} \bibnamefont{and}
  \bibinfo{author}{\bibfnamefont{X.}~\bibnamefont{Waintal}},
  \bibinfo{journal}{Phys. Rev. B} \textbf{\bibinfo{volume}{77}},
  \bibinfo{pages}{115119} (\bibinfo{year}{2008}).

\bibitem[{\citenamefont{Wurm et~al.}(2010)\citenamefont{Wurm, Wimmer, Baranger,
  and Richter}}]{Wurm2010a}
\bibinfo{author}{\bibfnamefont{J.}~\bibnamefont{Wurm}},
  \bibinfo{author}{\bibfnamefont{M.}~\bibnamefont{Wimmer}},
  \bibinfo{author}{\bibfnamefont{H.~U.} \bibnamefont{Baranger}},
  \bibnamefont{and} \bibinfo{author}{\bibfnamefont{K.}~\bibnamefont{Richter}},
  \bibinfo{journal}{Semiconductor Science and Technology}
  \textbf{\bibinfo{volume}{25}}, \bibinfo{pages}{034003}
  (\bibinfo{year}{2010}).

\bibitem[{\citenamefont{Chaves et~al.}(2009)\citenamefont{Chaves, Farias,
  Peeters, and Szafran}}]{Chaves2009a}
\bibinfo{author}{\bibfnamefont{A.}~\bibnamefont{Chaves}},
  \bibinfo{author}{\bibfnamefont{G.~A.} \bibnamefont{Farias}},
  \bibinfo{author}{\bibfnamefont{F.~M.} \bibnamefont{Peeters}},
  \bibnamefont{and} \bibinfo{author}{\bibfnamefont{B.}~\bibnamefont{Szafran}},
  \bibinfo{journal}{Phys. Rev. B} \textbf{\bibinfo{volume}{80}},
  \bibinfo{pages}{125331} (\bibinfo{year}{2009}).

\bibitem[{\citenamefont{Szafran and Peeters}(2005)}]{Szafran2005a}
\bibinfo{author}{\bibfnamefont{B.}~\bibnamefont{Szafran}} \bibnamefont{and}
  \bibinfo{author}{\bibfnamefont{F.~M.} \bibnamefont{Peeters}},
  \bibinfo{journal}{Phys. Rev. B} \textbf{\bibinfo{volume}{72}},
  \bibinfo{pages}{165301} (\bibinfo{year}{2005}).

\bibitem[{\citenamefont{B\"uttiker}(1986)}]{Buttiker1986a}
\bibinfo{author}{\bibfnamefont{M.}~\bibnamefont{B\"uttiker}},
  \bibinfo{journal}{Phys. Rev. Lett.} \textbf{\bibinfo{volume}{57}},
  \bibinfo{pages}{1761} (\bibinfo{year}{1986}).

\bibitem[{\citenamefont{Seelig and B\"uttiker}(2001)}]{Seelig2001a}
\bibinfo{author}{\bibfnamefont{G.}~\bibnamefont{Seelig}} \bibnamefont{and}
  \bibinfo{author}{\bibfnamefont{M.}~\bibnamefont{B\"uttiker}},
  \bibinfo{journal}{Phys. Rev. B} \textbf{\bibinfo{volume}{64}},
  \bibinfo{pages}{245313} (\bibinfo{year}{2001}).

\bibitem[{\citenamefont{Levkivskyi and Sukhorukov}(2008)}]{Levkivskyi2008a}
\bibinfo{author}{\bibfnamefont{I.~P.} \bibnamefont{Levkivskyi}}
  \bibnamefont{and} \bibinfo{author}{\bibfnamefont{E.~V.}
  \bibnamefont{Sukhorukov}}, \bibinfo{journal}{Phys. Rev. B}
  \textbf{\bibinfo{volume}{78}}, \bibinfo{pages}{045322}
  (\bibinfo{year}{2008}).

\bibitem[{\citenamefont{Kovrizhin and Chalker}(2010)}]{Kovrizhin2009a}
\bibinfo{author}{\bibfnamefont{D.~L.} \bibnamefont{Kovrizhin}}
  \bibnamefont{and} \bibinfo{author}{\bibfnamefont{J.~T.}
  \bibnamefont{Chalker}}, \bibinfo{journal}{Phys. Rev. B}
  \textbf{\bibinfo{volume}{81}}, \bibinfo{pages}{155318}
  (\bibinfo{year}{2010}).

\bibitem[{\citenamefont{Kotim\"aki and R\"as\"anen}(2010)}]{Kotimaki2010a}
\bibinfo{author}{\bibfnamefont{V.}~\bibnamefont{Kotim\"aki}} \bibnamefont{and}
  \bibinfo{author}{\bibfnamefont{E.}~\bibnamefont{R\"as\"anen}},
  \bibinfo{journal}{Phys. Rev. B} \textbf{\bibinfo{volume}{81}},
  \bibinfo{pages}{245316} (\bibinfo{year}{2010}).

\end{thebibliography}

\end{document}